\documentclass[12pt]{aastex}
\usepackage{emulateapj5,apjfonts,amsmath}
\usepackage{amsmath}
\newcommand{\bea}{\begin{eqnarray} }
\newcommand{\eea}{\end{eqnarray}}

\begin{document}
\submitted{Accepted for publication in ApJ, v. 591 (July 1, 2003)}
\title{Feedback From the First Supernovae in Protogalaxies: 
The Fate of the Generated Metals}

\author{Keiichi WADA$^{1,2}$, and Aparna VENKATESAN$^{2,3}$}
\altaffiltext{1}{National Astronomical Observatory of Japan, Mitaka, Tokyo
181-8588, Japan\\
E-mail: wada.keiichi@nao.ac.jp}
\altaffiltext{2}{CASA, UCB 389, University of Colorado, Boulder, CO
  80309-0389}
\altaffiltext{2}{NSF Astronomy and Astrophysics Postdoctoral Fellow}

\begin{abstract}
We investigate the chemo-dynamical effects of multiple supernova (SN)
explosions in the central region of 
primordial galaxies (e.g. $M\sim 10^8 M_\odot$ halo at $z \sim 10$)
using three-dimensional hydrodynamical
simulations of the inhomogenous interstellar medium down to parsec-scales.
We find that the final protogalactic structure and metal distribution
depend strongly on the number of SNe. Specifically, 1) 1000 SNe after an
instantaneous burst of star formation are sufficient to almost completely
blow away the gas in these systems, whereas 2) 100 SN explosions trigger
the collapse of the protogalactic cloud, leading to the formation of a
cold, dense clumpy disk ($n > 300$ cm$^{-3}$) with metallicity, $Z = 4
\times 10^{-4} Z_\odot$. 
These results imply that the metallicity of
the ``second generation'' of stars could be $Z \sim 10^{-4} Z_\odot$, and
that the environment to form metal-free stars in protogalaxies may be lost
relatively quickly ($\lesssim 10^7$ yr) after the first burst of $Z=0$
star formation.  The recently discovered ultra-metal-poor star (Christlieb
et al. 2002) may represent surviving members of such second-generation
star formation.
\end{abstract}


%
\section{INTRODUCTION}
%

The forefront problems of cosmology and star formation have become
increasingly intertwined in the last decade. This has been driven
especially by studies of the feedback from the first stars and supernovae
in high-redshift ($z$) protogalaxies on the intergalactic medium (IGM). The
collapse of primordial gas in dark matter (DM) halos followed by the
formation of the first metal-free stars has been extensively studied by
several groups (e.g. Omukai \& Nishi 1998; Abel, Bryan, \& Norman 2000;
Bromm, Coppi, \& Larson 2002).  Examining the feedback from the first
generation of supernovae (SNe) is essential for understanding the metal
enrichment and evolution of the interstellar medium (ISM), as well as the
structure and subsequent star formation in the host protogalaxies.  Some of
the key physical issues in this problem are the propagation of metal-rich
SN gas through the inhomogeneous ISM, and the subsequent mixing and
diffusion of metals in the multi-phase ISM.  Mori, Ferrara, \& Madau (2002;
hereafter MFM02), using three-dimensional hydrodynamical simulations, found
that a significant fraction of the metal-enriched gas in a $10^8 M_\odot$
halo at $z=9$ is blown away, and that multiple SN explosions in primordial
galaxies can play an important role in the metal pollution of the high-$z$
IGM.  However, the early stages of the interaction between multiple blast
waves from SNe in an inhomogeneous ISM are important for the subsequent
evolution of the ISM in protogalaxies. From this point of view, the minimum
resolved scale in MFM02-- 22 pc-- is not fine enough to resolve the
interactions between the blast waves and the clumpy ISM.  Furthermore,
self-gravity of the gas and radiative cooling below $10^4$ K, both of which
are not considered in MFM02, are crucial factors in determining the
internal structure of primordial galaxies after the initial SN
explosions. This is especially relevant for cases in which the initial
starburst is not strong enough to blow away the protogalactic gas.

In this paper, we report on the effects of an initial metal-free starburst
in primordial galaxies, using three-dimensional hydrodynamical simulations
with spatial resolution 3.9 pc, which take into account the self-gravity of
the gas and radiative cooling between 20 K and $10^8$ K. We focus here on
the simplest scenario, in which all the SNe are coeval and spatially
concentrated; variations on these initial conditions will be considered in
our future work.  We explore feedback from relatively small numbers of SNe,
in order to see what kind of ISM structure and metal distribution are
achieved in primordial galaxies just after the {\it first} SN explosions of
first-generation stars.  Studying SN feedback at these ISM scales has
important consequences not only for the metal pollution of the high-$z$
IGM, but also for the duration of metal-free star formation in the early
universe.

%
\section{NUMERICAL METHODS AND MODELS}
%
The hydrodynamical schemes which we use are described in Wada \&
Norman (2001) and Wada (2001). 
We follow explicitly the diffusion of metals ejected from SNe into the ISM by
solving the equation similar to that of mass conservation and assuming
that the metal component has the same local velocity as the fluid.
We calculate the gas cooling rate using cooling functions (Spaans, private
communication) for $20 < T_g < 10^8 {\rm K}$ 
and for gas metallicities in the range $Z = 0-1Z_\odot$.
In the primordial gas, $\rm{H}_2$ is the main coolant below $10^3$ K.
The $\rm{H}_2$ and HD abundances are determined through a chemical network. 
The diffusion of metals ejected from each SN into the ISM is explicitly
followed.  Therefore the gas metallicity is a function of position and
time.  The radiative cooling for the gas in each grid cell is determined
for the local metallicity \citep{gold78,suther93,spaans96}.  Note that in
reality the CMB sets a physical temperature floor for the radiative
cooling, e.g. $\sim 30$ K at $z \sim 10$ (Bromm et al. 2002), although the
gas temperature below $\sim 100$ K is not very important for the evolution
of the blast waves.

The hydrodynamic part of the basic equations is solved by AUSM (Advection
Upstream Splitting Method).
We used $512^2 \times 128$ equally spaced Cartesian grid points
covering a volume of 2 kpc $\times $ 2 kpc $\times$ 1 kpc. Therefore, the
spatial resolution is 3.9 pc and 7.8 pc for the $x-y$ and $z$ directions
respectively.
The Poisson equation is solved to calculate the self-gravity of the gas
using the FFT.  The second-order
leap-frog method is used for the time integration.  We adopt the implicit
time integration for the cooling term.

To achieve density and temperature distribution in a dark
matter potential for the initial condition, we performed a {\it
pre}-calculation in which a geometrically thin, dense disk is evolved in
the spherically symmetric, time-independent DM potential, $\Phi_{\rm DM}$.
We assumed $\Phi_{\rm DM} \equiv -(27/4)^{1/2}v_c^2/(r^2+ a^2)^{1/2}$,
where $a = 267$ pc is the core radius of the potential, and $v_c = 11$ km
s$^{-1}$ is the maximum rotational velocity.  The mass of the dark matter
in the central part of the protogalaxy is therefore $2 \times 10^7
M_\odot$.  The total mass and initial temperature of the gas are $10^7
M_\odot$ and 20 K.  The gas is uniformly distributed in a thin axisymmetric
disk, whose radius and scale height are respectively 400 pc and 8 pc.  The
total DM mass of $\sim 10^8 M_\odot$ corresponds to a virial radius of
$\sim 1 $ kpc at $z \sim 10$ (MFM02; Madau, Ferrara, \& Rees 2001).  Note
that the very first stars could form in smaller systems of mass $\sim
10^{5-6} M_\odot$ at higher redshift ($z\sim 20-30$) as studied in Bromm,
Coppi, \& Larson (2002).  Moreover, in the low-mass galaxies considered
here, star formation followed by SNe might already have occurred during the
initial collapse of the DM component.  For simplicity, and owing to the
short dynamical time scale of the gas component after the SN explosions, we
assume here that the DM potential is time-independent. A more
self-consistent treatment will be considered in future work.


In the upper panels of Fig. 1 ($t=0$), the density and temperature
distribution of the gas after virialization are shown.  During the
virialization process, the density distribution becomes almost spherical,
although some dense clumps are formed in the central region where the
temperature is less than $10^3$ K. The mean density and temperature within
100 pc from the center are 3 cm$^{-3}$ and 100 K.  The total gas mass in
the computing box is 5.6 $\times 10^6 M_\odot$. Gas in the cold phase ($T_g
< 100$ K) occupies about 1.4\% of the volume, and has a mass of 1.9 $\times
10^6 M_\odot$.  The maximum density in the dense clumps is $20$ cm$^{-3}$
(Jeans length $\sim 18$ pc).  We use this density, temperature, and
velocity distribution as the initial condition.

In order to assess the chemo-dynamical effects of the first SNe, some
assumptions must be made about the initial mass function (IMF) of the first
stars. There are significant uncertainties associated with this, since the
processes that determine the stellar IMF, even in the local universe, are
at present unclear.  Some recent theoretical studies (e.g., Abel, Bryan, \&
Norman 2000; Bromm, Coppi, \& Larson 2002) indicate that the primordial IMF
may have been weighted towards masses exceeding 100 M$_\odot$. If this were
true, the metal yield per SN could potentially be much higher than that for
the ``usual'' Type II SN. However, the duration and universality of the
processes that lead to such an IMF are currently unknown.  Given the
unresolved nature of this topic, we assume a present-day Salpeter IMF over
a stellar mass range of $1-100 M_\odot$ in this work.  We include metal
generation only from $Z=0$ stars of mass $8-40 M_\odot$ (Woosley \& Weaver
1995), assuming that stars more massive than 40 $M_\odot$ implode directly
into a black hole without mass/metal ejection into the ISM. For these
assumptions, each unit mass of gas that forms stars produces total
IMF-integrated masses in metals and stellar ejecta of 0.00663 and 0.1863
after the SN explosion. Furthermore, for the above assumptions, there is
one SN for every $\sim$ 100 $M_\odot$ of stars formed.  Thus, the total
metal yield and ejecta mass per SN on average are 0.663 $M_\odot$ and 18.63
$M_\odot$.  We input the energy ($10^{51}$ erg per SN) as thermal energy
and mass/metal yields from the SNe at time $t=0.$ We also explored cases in
which the SNe are not coeval, and occur over a period of $\sim 1$
Myr. However, no essential differences from the models discussed in this
paper were found.

Since the IMF and star formation efficiency in primordial galaxies are
poorly understood, we take the number of SNe, $N_{\rm SN}$, as a free
parameter, and we explore two models here: model A ($N_{\rm SN} = 10^3$)
and model B ($N_{\rm SN} = 10^2$).  We put in $N_{\rm SN}$ SNe randomly
within the central 100 pc, since some of the initial star formation is
expected in the dense, cold clumps seen in Fig. 1 ($t=0$).  The energy and
ejected masses in metals and gas from each SN are injected into a randomly
selected grid cell, around which a blast wave is then expanded.  This wave
is not necessarily spherical, owing to the inhomogeneous density field and
interactions with other SNe.

A priori, we expect that if the total input energy is sufficiently larger
than the binding energy, then the blast waves will blow the gas away, as
has been predicted in many papers (e.g. Larson 1974, and references above).
On the other hand, if the total SN energy is not efficiently converted into
the ISM kinetic energy, the multiple blast waves would form a cavity in the
protogalactic cloud.  We would then expect the whole system to become
dynamically unstable and collapse. This can be demonstrated in an analytic
estimate as follows.  Using a similarity solution of a blast wave in a
radial density profile going as, $\rho \propto r^{-2}$, the maximum radius
of blast waves, $r_s = (6 E/\bar{\rho})^{1/5} \tau_c^{2/5}$ for $\gamma =
5/3$ (Ostriker \& McKee 1988), where $E$ is the input energy, $\bar{\rho}$
is the average density, and $\tau_c$ is the cooling time of the hot gas in
the cavity generated by the SNe.  In our work here, $\tau_c \sim 2 $ Myr,
so that the cavity radius would be $\sim$ 200 pc $(\varepsilon/0.1)^{1/5}
(N_{\rm SN}/100)^{1/5} (\bar{\rho}/0.03 M_\odot {\rm pc}^{-3})^{-1/5}
(\tau_c/2 {\rm Myr})^{2/5}$, where $E = \varepsilon N_{\rm SN} 10^{51}$
erg, and $\varepsilon$ is the heating efficiency (the fraction of the total
SN energy that is converted to the kinetic energy of the blast waves).
When the hot gas in the cavity cools, we expect the protogalactic gas
outside the cavity to lose pressure support and to begin falling towards
the galactic center. The gas mass inside the cavity ($r< r_s=200 {\rm pc}$)
is about $10^6 M_\odot$. If all the gas in the protogalaxy, about $5.6
\times 10^6 M_\odot$, were to collapse and mix uniformly with the hot
metal-enriched cavity, then the resulting average gas metallicity would be
about 100 SN $\times$ (0.7 $M_\odot$/SN)/(6.6 $\times 10^6 M_\odot$) $\sim
5\times 10^{-4} Z_\odot$.

In reality, the above estimates will be affected by the evolution of the
multiple blast waves in an inhomogeneous medium, the fraction of
protogalactic gas that eventually collapses, and by the heating efficiency,
$\varepsilon$, which in this situation is unknown.  In addition, the mixing
of metals in the collapsing protogalaxy will be nonuniform.  In our
numerical method here, we do not have to make any specific assumptions for
these factors. Instead, we explicitly solve for the metal diffusion, and
the interaction between the multiple blast waves and the inhomogeneous ISM.

%
\section{RESULTS}
%

The evolution of the density and temperature distributions for model A are
shown in Figure 1. The multiple SN explosions evolve as one big blastwave,
which reaches the outer boundary of the $z-$direction at $t=2.1$ Myr, and
those in the $x-$ and $y-$ directions at $t\sim 6$ Myr.  As seen in the
figures, the explosions are roughly spherical, but dense `spikes' are
formed, caused by non-linear instabilities in the decelerated shocks
\citep{yoshida92}.  The hot gases in the cavity quickly cool from $\sim
10^7$ K at $t=2.1$ Myr to $\sim 10^3$--$10^6$ K at $t=5.8$ Myr.  At $t = 3$
Myr, the gas velocity is 2-4 times larger than the escape velocity in the
whole region.  This implies that 1000 SNe are sufficient to blow away the
gas in the protogalaxy in $\sim 10^7$ Myr.

On the other hand, in model B, the protogalaxy survives the initial
starburst. The 100 SNe generate a cavity in the central region of the
protogalaxy, but it does not exceed sizes of $r\sim 300$ pc.  After $t\sim
20$ Myr, the cavity collapses, which in turn triggers the collapse of the
protogalactic gas cloud. Eventually a clumpy disk-like structure, whose
radius is about 200 pc, is formed in the center (Fig. 2).  The $y-z$ cross
section reveals a `chimney'-like structure.  The diffuse gases seen in
Fig. 2 are still falling in towards the central dense disk.  At $t=52.6$
Myr, the mass of dense gas ($n >$ 30 and 300 cm$^{-3}$) is 1.6$\times 10^6
M_\odot$ and 7.2$\times 10^5 M_\odot$, which are 36\% and 16\% respectively
of the total gas mass in the box at that time. The maximum density is
$3.4\times 10^4$ cm$^{-3}$.

Fig. 2 (c) shows that the metallicity distribution in the `chimney' is
about ($10^{-3}$--$10^{-2}$) $Z_\odot$, and that the diffuse halo is almost
metal-free.  The average metallicities in the very dense regions of $n >$
30 and 300 cm$^{-3}$ are 6.5$\times 10^{-4} Z_\odot$ and 4.1$\times 10^{-4}
Z_\odot$. These are in contrast to $Z > 10^{-2}Z_\odot$ in the relatively 
low density regions between the dense clumps in the central disky structure, owing to the
formation of these clumps through the accretion of mainly metal-poor or
metal-free gas.

As one may expect, the two models have different consequences for the
global ejection of metals.  Fig. 3 shows the evolution of the radial
metal mass distribution with time in the two models. 
Almost all the metals ($6\times 10^2 M_\odot$) in model A
are ejected out of the central protogalactic region, whereas most of the
metals in model B return to $r < 200$ pc in 50 Myr.  As a result, the
central region in model B is ten times more metal-rich than in model A.

In addition, the efficiency of converting the SN energy to the kinetic
energy of the ISM is time-dependent. In model B, this efficiency, defined
as the ratio of the total ISM kinetic energy to total input energy, is 0.12
at $t=1$ Myr, and $7\times 10^{-3}$ at $t=10$ Myr.

In summary, our results in the context of the simplified models we have
considered here-- SN feedback from an instantaneous starburst in
protogalaxies of mass about $10^7$--$10^8 M_\odot$-- are as follows. First,
protogalaxies which undergo bursts with more than 10$^3$ SNe are not likely
to retain the generated metals.  The ultimate fate of the ejected gas and
metals in this case is unclear, as they have not been tracked for
sufficiently long timescales here. They may eventually escape to the IGM,
or fall back partially to the galactic center, forming a dense core as
found in the large dynamic range simulations by MFM02.  Second, we find
that less active starbursts ($N_{\rm SN} \sim 100$) could lead to faster
metal re-incorporation in the gas, from which ``second generation'' stars
with $Z \sim 10^{-4} Z_\odot$ could potentially form.  The value of this
transition metallicity is indicative rather than exact, given that we do
not explicitly track the process of star formation itself, and that mixing
processes depend on instabilities that may not be entirely resolved by our
simulations.  Nevertheless, this metallicity is interesting in terms of the
recent discovery of an ultra-metal-poor star with [Fe/H]$=-5.3$ (Christlieb
et al. 2002).  The abundance pattern of elements in this object suggests
that it may have formed from a gas cloud that was pre-enriched by Type II
SNe. Such a star could represent the surviving members of the
second-generation stellar population in our scenario's model B, but with
ten times fewer initial SNe. In addition, our derived value for the
transition metallicity is in agreement with the numerical estimates in
recent works (Bromm et al. 2001; Mackey, Bromm, \& Hernquist 2003;
\citealt{schneider02}), for the critical level of enrichment that must
occur in order for star formation in primordial gas to evolve from a
hypothesized top-heavy stellar IMF to a present-day IMF. Future
observations of ultra-low-$Z$ stars will strongly test these theoretical
predictions for the metallicity and masses of second generation stars.



%
\section{DISCUSSION}
%

In this paper, we have begun an investigation of SN feedback in high-$z$
protogalaxies. Our methods contain significant improvements over recent
works, such as the inclusion of gas self-gravity in 3D and improved spatial
resolution. These are necessary to quantitatively understand the effects of
multiple SNe from the first stars at scales which are important for the
ISM, which influences the cosmological relevance of early SNe for the
IGM. We find that the fate of the generated metals-- whether they are
transported as a hot gas phase to the IGM or reincorporated quickly into
cold starforming gas in the protogalaxy-- depends critically on the number
of initial SNe. This has broad implications for the transition metallicity
at early epochs from metal-free to ``second generation'' star formation,
and the rapidity of this process. The duration of metal-free star formation
is of critical importance to theoretical studies of the H~I and He~II
reionization of the IGM (see, e.g., \citealt{venk03}), as well as for
observational searches for such stars in the Galactic halo or at high
redshift (\citealt{tum03}, and references therein).

In the initial condition, the total mass of cold ($T_g < 100$ K), dense gas
($n > 3$ cm$^{-3}$) is $3\times 10^5 M_\odot$. The gas mass converted to
stars in model A, with $10^3$ SNe, is about $10^5 M_\odot $.  Therefore, if
the star formation efficiency is larger than $\sim 0.3$, the first objects
formed in the protogalaxies are almost pure stellar systems of mass $\sim
10^5 M_\odot$, and the remaining baryonic mass is ejected from the galaxy.
These metal-free stellar systems would not be bound, because a large
fraction of the initial gas mass in the protogalaxy's central region is
lost due to strong winds from the starburst's SNe.  On the other hand, if
the initial star formation efficiency is of order a few percent, the first
starburst does not inhibit the subsequent creation of new star formation
sites.  If the star formation efficiency remains low, the metals from SNe
cannot be expelled from the protogalaxies. In this case, metal-enriched
stellar clusters of mass $\sim 10^{5}-10^{6} M_\odot$ eventually form,
accompanied by a diffuse gaseous halo.  These systems can be the building
blocks of larger galaxies, or the progenitors of globular clusters.  In
other words, suppose that the star formation efficiency is constant,
e.g. 0.1, among protogalaxies. The smaller systems ($M_{\rm total} < 10^7
M_\odot$) would then dominate the metal enrichment of the IGM, which is
consistent with the findings of Mac Low \& Ferrara (1999), and Ricotti,
Gnedin \& Shull (2002). The fate of the metals generated by the first stars
is clearly linked closely to the mass and star formation histories of the
host protogalaxies. In this work, we have considered a simple burst
scenario in which the SNe were coeval and spatially concentrated, in order
to focus on the ISM gas dynamics. We will examine the importance of varying
these and other factors, such as the stellar IMF, in more detail in the
future.

%
\acknowledgments 
%
We are grateful to Andrea Ferrara and Massimo Ricotti for their useful
comments, and for discussions on the simulations' initial condition.  We
thank the anonymous referees for helpful suggestions that improved the
manuscript, and Marco Spaans for providing the gas cooling functions.
Numerical computations were carried out on Fujitsu VPP5000 at NAOJ.
A.~V. gratefully acknowledges the support of NSF grant AST-0201670.


\clearpage
\begin{figure}
\caption{Log-scaled density ($M_\odot$ pc$^{-3}$) and temperature (K)
distributions of the initial condition ($t=0$), and at two snapshots
($t=$2.1 Myr and 5.8 Myr) of model A ($N_{\rm SN} = 10^3$).  (a) The $x-y$
cross sections of density, (b) the $y-z$ cross sections of density, and (c) same as (b), but for temperature.}
\end{figure}

\begin{figure}
\caption{Same as Figure 1, but for model B ($N_{\rm SN} = 10^2$) at $t=$
56.2 Myr. The lower right panel shows the $y-z$ cross section of the
metallicity distribution in the central region. The plotted range of
metallicities is $Z = 10^{-4}Z_\odot$--$10^{-1}Z_\odot$. }
\end{figure}

\begin{figure}
      \epsscale{0.6}
\plotone{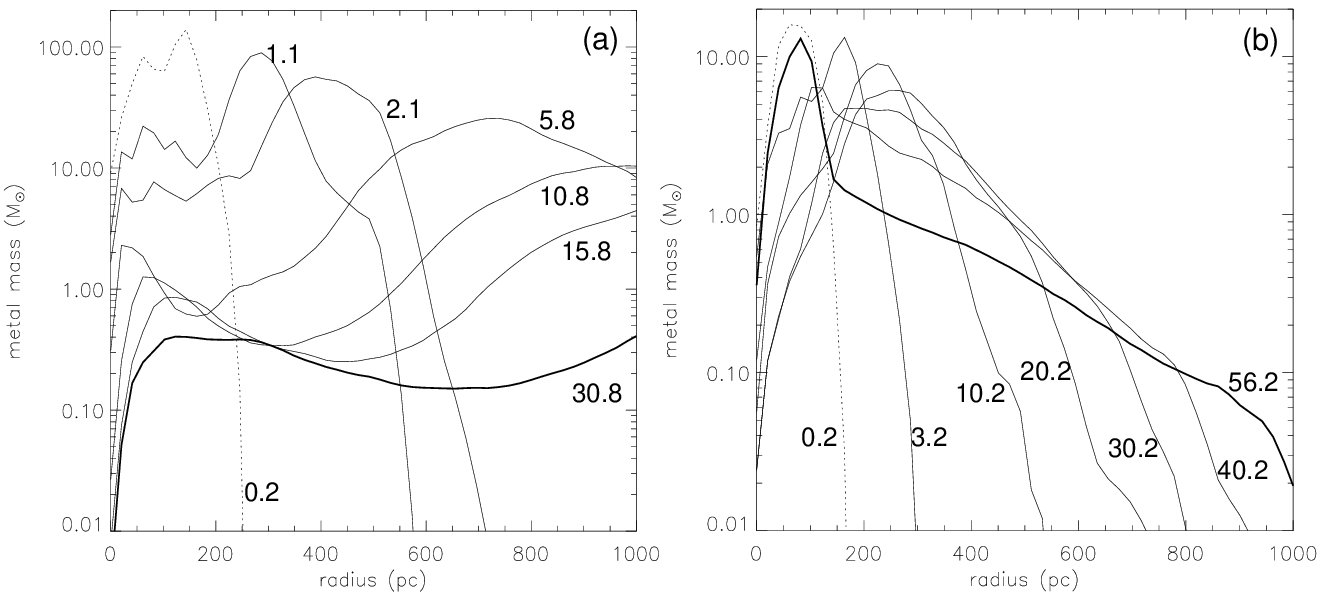}
\caption{The time evolution of the radial distribution of metals in (a)
model A and (b) model B.  The individual curves are labelled with time in
Myr. The metal mass is averaged over volume shells of width 20 pc at each radius.}
\end{figure}


\begin{thebibliography}{}
\bibitem[Abel, Bryan, \& Norman(2000)]{abel00} Abel, T., Bryan G. L., \& Norman, M. L. 2000, ApJ, 540, 39
\bibitem[Christlieb et al.(2002)]{cris02} Christlieb, N., Bessell, M. S., Beers, T. C., 
Gustafsson, B., Korn, A., Barklem, P. S., Karlsson, T., Mizuno-Wiedner, M., \& Rossi, S. 2002, Nature, 419, 904.
\bibitem[Bromm, Coppi, \& Larson(2002)]{bromm02} Bromm, V., Coppi, P.~S., \& Larson, R.~B.\ 2002, \apj, 564, 23 
\bibitem[Bromm, Ferrara, Coppi, \& Larson(2001)]{bromm01} 
Bromm, V., Ferrara, A., Coppi, P.~S., \& Larson, R.~B.\ 2001, \mnras, 328, 
969 
\bibitem[Goldsmith \& Langer(1978)]{gold78} Goldsmith, 
P.~F.~\& Langer, W.~D.\ 1978, \apj, 222, 881 
\bibitem[Larson(1974)]{larson74} Larson, R.~B.\ 1974, \mnras, 
169, 229 
\bibitem[Mac Low \& Ferrara(1999)]{mac99} Mac Low, M.~\& Ferrara, A.\ 1999, \apj, 513, 142
\bibitem[Madau, Ferrara, \& Rees(2001)]{madau01} Madau, P., Ferrara, A., \& Rees, M.~J.\ 2001, \apj, 555, 92 
\bibitem[Mori, Ferrara, \& Madau(2002)]{mori02} Mori, M., Ferrara, A., \& Madau, P.\ 2002, \apj, 571, 40 (MFM02)
\bibitem[Mackey, Bromm, \& Hernquist(2003)]{mackey03} Mackey, J., Bromm, V., \&  Hernquist, L. 2003, ApJ, in press (astro-ph/0208447)
\bibitem[Omukai, \& Nishi(1998)]{omukai98} Omukai, K., \& Nishi, R. 1998, ApJ, 508, 141
\bibitem[Ostriker \& McKee(1988)]{ost88} Ostriker, J.~P.~\& McKee, C.~F.\ 1988, Reviews of Modern Physics, 60, 1 
\bibitem[Ricotti, Gnedin, \& Shull(2002)]{ricotti02} Ricotti, M., Gnedin, N.~Y., \& Shull, J.~M.\ 2002, \apj, 575, 49 
\bibitem[Schneider, et al.(2002)]{schneider02} Schneider, R., Ferrara, A., Natarajan, P., \& Omukai, K.\ 2002, \apj, 571, 30 
\bibitem[Spaans(1996)]{spaans96} Spaans, M.\ 1996, \aap, 307, 271 
\bibitem[Sutherland \& Dopita(1993)]{suther93} Sutherland, 
R.~S.~\& Dopita, M.~A.\ 1993, \apjs, 88, 253 
\bibitem[Tumlinson, Shull, \&  Venkatesan(2003)]{tum03}
{Tumlinson}, J., {Shull}, J.~M., \& {Venkatesan}, A. 2003, \apj, 584, 608
\bibitem[Venkatesan, Tumlinson, \&  Shull(2003)]{venk03}
{Venkatesan}, A., {Tumlinson}, J., \& {Shull}, J.~M. 2003, \apj, 584, 621
\bibitem[Wada(2001)]{wada01} Wada, K. 2001, \apjl, 559, L41 
\bibitem[Wada \& Norman(2001)]{wad01b} Wada, K.\ \& Norman, C.\ A.\ 2001, \apj, 547, 172  
\bibitem[Woosley \& Weaver(1995)]{woo95} Woosley, S.~E.~\& Weaver, T.~A.\ 1995, \apjs, 101, 181 
\bibitem[Yoshida \& Habe(1992)]{yoshida92} Yoshida, T.~\& Habe, 
A.\ 1992, Progress of Theoretical Physics, 88, 251 
\end{thebibliography}
\end{document}